\documentclass[a4paper,12pt]{article}

\usepackage{graphicx}  
\usepackage{url}  
\usepackage{float}  
\usepackage{xcolor}
\newcommand{\text}[1]{{\hbox{#1}}}
\newcommand{\R}{{\mathbb R}}
\newcommand{\C}{{\mathbb C}}

\newcommand{\Z}{{\mathbb Z}}
\newcommand{\Q}{{\mathbb Q}}

\usepackage[utf8]{inputenc}
\usepackage{amsmath}
\usepackage{amsfonts}
\usepackage{amssymb}
\usepackage{amsthm}

\usepackage{graphicx}
\usepackage{xcolor}
\usepackage{url}

\usepackage{latexsym}	
\usepackage{amsmath}
\usepackage{amsthm}
 \usepackage{url}

\usepackage{xcolor}

\usepackage{ amssymb }
\def\ba{\begin{eqnarray}}
\def\ea{\end{eqnarray}}

\def\H{\hbox{\bf H}}

\def\R{\hbox{\bf R}}

\def\ba{\begin{eqnarray}}
\def\ea{\end{eqnarray}}

\def\H{\hbox{\bf H}}

\def\R{\hbox{\bf R}}

\def\be{\begin{equation}}
\def\ee{\end{equation}}

\def\C{\mathbb C}
\theoremstyle{plain}

\def\C{\mathbb C}
\def\R{\mathbb R}

\begin{document}

\title{About the quantum Talbot effect on the sphere}

\author{Fernando Chamizo\thanks{Departamento de Matem\'aticas and ICMAT. Universidad Aut\'onoma de Madrid. 28049 Madrid, Spain,
fernando.chamizo@uam.es.} \and Osvaldo P. Santill\'an\thanks{Instituto de Matem\'atica Luis Santal\'o (IMAS), UBA CONICET, Buenos Aires, Argentina firenzecita@hotmail.com and osantil@dm.uba.ar.}}
\date{}
\maketitle

\begin{abstract}
The Schr\"odinger equation on a circle with an initially localized profile of the wave function 
is known to give rise to revivals or replications, where the probability density of the particle
is partially reproduced at rational times. 
As a consequence of the convolutional form of the general solution it is deduced that
a piecewise constant initial wave function remains
piecewise constant at rational times as well. For a sphere instead, 
it is known that this piecewise revival does not necessarily occur, indeed
the wave function becomes
singular at some specific locations at rational times. It may be desirable to study the same problem, but with an initial condition being a localized
Dirac delta instead of a piecewise constant function, and this is the purpose of the present work. By use of certain summation
formulas for the Legendre polynomials together with properties of Gaussian sums, it is found that revivals on the sphere
occur at rational times for some specific locations, and the structure of singularities of the resulting wave function is characterized
in detail. In addition, a partial study of  the regions where the density vanishes, named before valley of shadows in the context of the circle, is initiated here.
It is suggested that, differently from the circle case, these regions are not lines but instead some specific  set of points along the sphere. 
A conjecture about the precise form of this set is stated and the intuition behind it is clarified.
\end{abstract}

%
\vspace{2pc}
\noindent{\it Keywords}: Talbot effect, quantum revivals, Gaussian sums

\section{Introduction}

The Talbot effect is an image replication phenomena which has its origins in optics. The mathematical characterization of this effect in terms
of Fresnel diffraction leads to a wider characterization of this replication or revival phenomena in physics, which ranges in its application from quantum mechanics to statistical physics
and solid state scenarios.
This revival phenomena is characteristic of dispersive equations.

\subsection{Origins in optics}

In 1836, Sir M. F. Talbot observed that, when coherent light passes through a diffraction grating, the diffraction pattern  at integral multiples of certain distance replicates the structure of the grating. This phenomena was reported in a purely descriptive paper in \cite{talbot}. Later on, this effect was explained theoretically by Lord Rayleigh \cite{rayleigh} in 1881, and it was shown that it is a natural consequence of the Fresnel diffraction formulas. The diagram of the diffraction wave intensity at different distances of the grating has a fractal-like structure known as \emph{Talbot carpet}. The theoretical achievement of Lord Rayleigh is to give
an explicit expression $d_n=n\alpha^2/\lambda$, with $n$ an integer, for the distances from the grating where the image revival occurs. 
Here $\lambda$ is the wavelength of the light, assumed to be monochromatic, and $\alpha$ a constant describing the spacing of the grating (see below). In addition to the reconstruction images, several other images can be seen, which are known in the optics terminology as Fresnel images. These last ones are approximations, that is, they are not exact replicas of the initial object.

To build a simple mathematical, but not completely precise, intuition of the situation,
consider the slits in the diffraction grating  as an infinite set of point like holes at integers spaced values $y=nd$ of the $Y$-axis. If a monochromatic plane wave of frequency $\nu$ and wavelength $\lambda\ll 1$ enters the diffraction grating from the left half plane, each hole acts as a source of spherical waves. By employing the abbreviation $e(x)=e^{2\pi ix}$, we can model the diffracted wave far enough from the holes as a sum of cylindrical waves of the form
$$
 u(\mathbf{x},t)=
 \sum_{\mathbf{k}} \frac{e\big(|\mathbf{k}|r_n-\nu t\big)}{\sqrt{r_n}}
 \qquad\text{with}\quad
 |\mathbf{k}|=\lambda^{-1}\text{ and }k_2\in\Z.
$$
Here $\mathbf{k}$ is the wave vector as used in crystallography and it differs in a factor $2\pi$ from the standard one in physics. The quantity $r_n$ denotes the distance
of the observation point to the slit located at the position $n$. The square of the absolute value of this quantity is related to the intensity, which is explicitly given at distance $r=\sqrt{x^2+y^2}$ by $I(r)$ with
$$
 I(r)=\Big|\sum_n \frac{\cos\bigg(\frac{2\pi}{\lambda}\sqrt{x^2+(y-nd)^2}\bigg)}{\sqrt{r_n}}\Big|^2
$$
$$
+\Big|\sum_n \frac{\sin\bigg(\frac{2\pi}{\lambda}\sqrt{x^2+(y-nd)^2}\bigg)}{\sqrt{r_n}}\Big|^2.
$$
At $y=0$ the maximum intensity is found by requiring that the cosine terms are all equal to one. This implies that
$$
\frac{2\pi}{\lambda}\sqrt{x^2+(nd)^2}=m \pi.
$$
Assume for a moment that the large distance condition $(nd/x)^2<<1$ is satisfied. Then by Taylor expanding the square root it is found that
$$
\frac{2x}{\lambda}-\frac{n^2 d^2}{x\lambda}\sim m.
$$
As a simplifying assumption consider the case when $2x=k \lambda$ with $k$ an integer. If in addition $x\lambda=d^2$ then the last equation is satisfied for all $n$.
In other words, the position $x=d^2/\lambda$ is a maximum for all the terms.
This is the position found by Lord Rayleigh \cite{rayleigh} where the replication takes place. For $x=2d^2/\lambda$
several but not all the holes have a maximum, the resulting replication becomes partial, and the replication becomes worse for $x=l d^2/\lambda$ with $l$ an integer such
that $l>2$. In addition for $x=d^2/l \lambda$  the Talbot effect may appear at fractional distances, an issue which was
studied in \cite{BeKl}. However, if the distance becomes small, then the large distance approximation becomes not completely reliable. The analysis of this paragraph
is suggestive, although not completely rigorous, and we refer the reader to the original references for further details. 
{See, for instance, the classic review \cite{patorski}.}

In the above description the grating is composed by simple holes. 
It is more realistic to assign a certain width~$w$ to each slit and then the waves in the preceding sum are damped by the Fourier transform of the characteristic function of an interval (the sinc function), according to Frauenhofer diffraction theory. Then, we represent the diffraction intensity pattern by the time independent formula
$$
 |u(x,y,t)|^2=
 \bigg|
 \sum_{|n|\le \lambda^{-1}}
 \frac{\sin(n\pi w)}{n\pi}
 e\big(x\sqrt{\lambda^{-2}-n^2}+ny\big)
 \bigg|^2.
$$
It seems natural to assume that the biggest contribution to the sum corresponds to greater amplitudes, or equivalently, to smaller values of $n$. Under this assumption, the mathematical explanation of the replication of the structure of the grating in the Talbot effect is very simple. It reduces to 
$$
 \sqrt{\lambda^{-2}-n^2} -\lambda^{-1}\sim \frac{\lambda}{2}n^2,
$$
which fits in the \emph{paraxial approximation} of the geometrical optics.
Then the expression 
$$u(x,y,t)e(-\lambda^{-1} x),$$ is expected to be near invariant by $x\mapsto x+2\lambda^{-1}$ and we see a periodicity in the diffraction pattern. If we consider fractional multiples of this \emph{Talbot length} $2\lambda^{-1}$ then we have to take into account the cancellation encoded in the quadratic Gauss sums \cite{BeKl}. More technical details of the optical setting of this effect may be found in \cite{fresnel1}, \cite{fresnel2}, \cite{fresnel3}.
\smallskip

\begin{figure}[h!]
\centering
\includegraphics[width=0.5\textwidth]{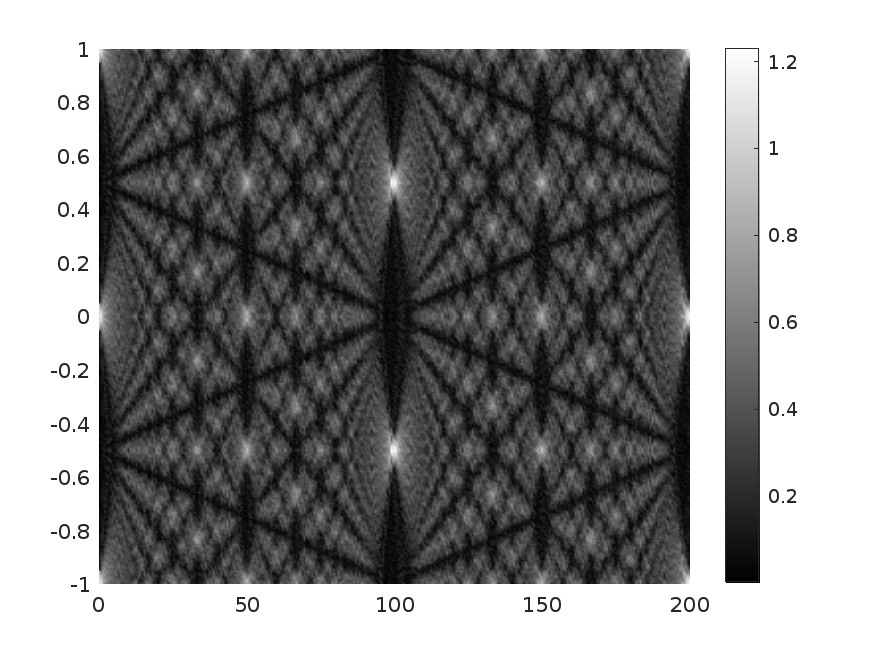}
\caption{Simulation of an optical Talbot carpet with $\lambda= 0.01$ and $w=0.1$.}
 \label{fig:1}
\end{figure}

As a final aside, although the Talbot effect is very old as an optical phenomenon,  it is not so simple to observe clearly.
In \cite{bakman} is described how to witness the Talbot carpet in a classroom experiment with water waves. On the other hand, it is fairly simple to write a computer program to simulate numerically Talbot effect with the preceding exponential sum\footnote{In \url{http://matematicas.uam.es/~fernando.chamizo/dark/d_talbot.html}  there are
some Talbot carpet simulations and code to generate them {and in \url{https://demonstrations.wolfram.com/TheTalbotCarpetInTheCausalInterpretationOfQuantumMechanics/} there is an interactive applet.}}. For instance, 
Figure~\ref{fig:1}
was obtained as the density plot of $|u(x,y,t)|$ for $\lambda^{-1}=100$ and $w=0.1$ using $300\times 300$ values of $(x,y)$.
The width is the Talbot length and we observe a perfect replication on the right side the three hot spots on the left representing three slits. They appear shifted at halfway  and  we see a copy scaled by a factor~$1/2$ at $1/4$ of the Talbot distance.
The Talbot effect has been studied for several different
grating profiles, and the replication of images is characteristic as well {(see \cite{montgomery} for conditions for this replication)}.

Even though the above discussion gives a general idea about the Talbot effect, it is of interest to describe it in the context of Fresnel diffraction. 
We make a brief discussion, and we again refer the reader to technical details to the original references. In a two dimensional setting, the Talbot effect is described by
the Helmhotz equation 
$$
\nabla^2 u+k^2 u=0,
$$
where the two dimensional Laplace operator is related to the coordinates $x$ and $y$, where $x$ is the distance from the grating and $y$ is the one dimensional coordinate
for the one dimensional wall (a line), where the holes are located.
The initial condition for point like holes is given by
\begin{equation}\label{unus}
u(0, y)=\sum_{n=-\infty}^{\infty} \delta(x-nd)=\frac{1}{d}\sum_{n=-\infty}^{\infty} e\bigg(\frac{ny}{d}\bigg).
\end{equation}
The solution with this initial condition is given by
$$
u(x, y)=\frac{1}{d}\sum_{n=-\infty}^{\infty} e\bigg(\frac{x}{\lambda} \sqrt{1-\bigg(\frac{dn}{\lambda}\bigg)^2}\bigg)e\bigg(\frac{ny}{d}\bigg).
$$
In the paraxial approximation described above, this solution may be approximated by
$$
u(x, y)= e\bigg(\frac{x}{\lambda} \bigg)v(x, y),
$$
where the following function was introduced
\begin{equation}\label{berro}
v(x,z)= \frac{1}{d}\sum_{n=-\infty}^{\infty} e\bigg(\frac{x}{2\lambda}\bigg(\frac{dn}{\lambda}\bigg)^2\bigg) e\bigg(\frac{ny}{d}\bigg).
\end{equation}
This function satisfies a Schr\"odinger like equation
$$
v_x=i c v_{yy},
$$
where $c$ is a constant depending on the lengths of the problem, and with the initial conditions \eqref{unus}. This motivates the study of the fractal properties of the Schr\"odinger equation, even in the context of optics. In the present case 
the role of the time is played by the coordinate $x$.

\subsection{Applications in general wave phenomena}

The description given about the Talbot effect suggests a connection with the Schr\"odinger equation, there came to light that such replication or revival phenomena is characteristic 
of linear hyperbolic differential equation. An example is of course the Schr\"odinger equation itself \footnote{From the mathematical point of view, the  Schr\"{o}dinger equation is not included in the classification due to the fact that its coefficients are not real. When it is included, it is classified as parabolic, not hyperbolic.}. In this case a single particle concentrated
initially in some location is concentrated again near some orbital numbers at certain specific times.  The Schr\"odinger equation is a particular 
case of dispersive equations \cite{ChOl}, \cite{olver}, in which the revival phenomena was found in form of dispersively quantized cusps at rational times in several contexts. 

The Talbot effect has attracted the attention in mathematics as well, mainly due to the conjecture in \cite{BeKl}, \cite{BeK2} about the fractal nature of the real and imaginary parts
of the function \eqref{berro} and its generalization for different gratings. Some mathematically founded results can be found in \cite{oskolkov}, \cite{eceizabarrena}, \cite{rigor1}, \cite{rigor2}, \cite{rigor3} and references therein. There is a connection to variations on the so called Riemann's example, a Fourier series that according to Weierstrass was proposed by Riemann as an example of continuous and nowhere differentiable function which has interesting fractal and self-similar properties \cite{duistermaat}, \cite{jaffard} \cite{chamizo}.

In addition, it was understood later on that the Talbot effect is a particular case of replication phenomena, with a wide application in physics. Replications
in the context of relativistic Dirac equations are more difficult to be observed, nevertheless some theoretical basis was provided in \cite{relativistic}.
The replication phenomena in Dirac equation in fact may have applications in the physics of graphene \cite{graphene}. There are theoretical works
about revival phenomena in conformal field theories \cite{cardy1}, \cite{cardy2}, in tight binding models \cite{tight}, Loschmidt echoes \cite{loschmidt}-\cite{pastawski4} or in quantum walks on cycles \cite{cycles}.
In addition, this effect was studied for non linear generalizations of the Schr\"odinger equation which describes the dynamics of vortex filaments in \cite{filamento1}, \cite{filamento2}. The connection with Riemann's example reappears here.

{A recent broad review considering the classic and the quantum settings and entering into the applications is \cite{Wen}.

\subsection{Brief outline of the present work}

The present work is aimed for the characterization of  the revivals of a quantum particle on a sphere, under the assumption that the initial wave function is a delta function.
Although there exists related work about this problem, namely \cite{HaLo}, the initial conditions that are considered differ from other works on the literature.
The goal is to clarify the structure of singularities of the wave function, which requires some specific summation formulas and general properties of Gauss sums.
In addition, some emphasis is placed on the classification of the shadow regions, that is, the spatial set for which the probability density vanish identically.

The present work is organized as follows. In section 2, the mathematical problem to be solved is stated. In section 3 and 4 some useful identities related to Legendre polynomials
and Gauss sums are reviewed, which are of importance in the subsequent analysis. In section 5 and 6 the solution is found explicitly for certain rational times, and the singular structure of this solution is explicitly characterized. In section 7 the notion of valley of shadows introduced in \cite{oskolkov} for the circle is reviewed, and a conjecture generalizing these statements
to the spherical case is presented. Section 8 shows a link of the presented results to the Talbot effect and section 9 contains a discussion of the results.

\section{The quantum setting}

The quantum version of the Talbot effect is related to the so called \emph{quantum revivals}. The Schr\"odinger equation is a kind of imaginary heat equation 
and then we may expect dispersion when the time goes by (as a manifestation of uncertainty). In some geometric situations however, we expect to recover, fully or approximately, the initial state.

We consider here the unit sphere $S^2$ as the base space. The Schr\"odinger equation on it, after adjusting the units, is
$$
 i\frac{\partial\Psi}{\partial t}
 =
 -\nabla^2_{S^2}\Psi
 \qquad\text{with}\quad
 \nabla^2_{S^2}
 =
 \frac{1}{\sin^2\theta}
 \Big(
 \big(\sin\theta \frac{\partial}{\partial\theta}\big)^2
 +\frac{\partial^2}{\partial\varphi^2}
 \Big).
$$
The eigenfunctions of $-\nabla^2_{S^2}$ are $e^{im\varphi}(\sin\theta)^{|m|} P_\ell^{(|m|)}(\cos\theta)$, with $-\ell\le m\le \ell$ and  $P_\ell$ the Legendre polynomials \cite[\S VII.5.3]{CoHi1}. The corresponding eigenvalues are $\ell(\ell +1)$ with multiplicity $2\ell+1$ due to the restriction $-\ell\le m\le \ell$. Then any polar solution (meaning not depending on the azimuthal angle) admits an expansion
\begin{equation}\label{ssol}
 \Psi(\theta,t)
 =
 \sum_{\ell=0}^\infty
 a_\ell P_\ell(\cos\theta)e^{-i\ell(\ell+1)t}.
\end{equation}
Note that $\ell(\ell+1)$ is always even, 
then $\Psi$ is $\pi$-periodic in time and we have a trivial integral Talbot effect, meaning that for $k\in\Z$ these solutions satisfy 
$\Psi(\theta,t)=\Psi(\theta,t+k\pi)$.

In 
\cite{HaLo} the singularities appearing in the fractional Talbot are studied under the initial condition $\Psi(\theta,0)=\textrm{sgn}(\pi/2-\theta)$. The choice of this particular example is not well motivated. In fact, in \cite{HaLo} we can read {\sl The most fundamental case to investigate would be the time evolution of a single point $\delta$-function initial wavefunction} [\dots] {\sl but this would be more involved} [\dots].
Here we take the challenge of studying the problem for an initial condition whose associated probability density becomes  the Dirac $\delta$ in the limiting case. More precisely, we are going to consider  a problem of the form
\begin{equation}\label{pde}
 i\frac{\partial\Psi}{\partial t}
 =
 -\nabla^2_{S^2}\Psi
 \qquad\text{with}\quad
 \Psi(\theta,0)=f_r(\theta)
\end{equation}
where $f_r$ is a certain regular function depending on the parameter $0<r<1$ and $|f_r|^2$ tend to the Dirac $\delta$ at the north pole of the sphere as $r\to 1^-$.

\section{An approximation of the identity}

Historically, the Legendre polynomials appeared as the coefficients in the multipole expansion of the Newtonian potential. Namely, if $r=|\mathbf{x}|< 1$ and $|\mathbf{y}|=1$ we have \cite[\S VII.5.5]{CoHi1}
$$
 \frac{1}{|\mathbf{x}-\mathbf{y}|}
 =
 \sum_{\ell=0}^\infty
 r^\ell P_\ell(\cos\theta)
 \qquad\text{with}\quad
 \theta=
 \measuredangle \mathbf{x},\mathbf{y}.
$$
Indeed this formula can be related to the spectral properties of $P_\ell$ \cite[\S8.4]{taylor2}.
If $\mathbf{x}$ is on the $z$-axis then $\theta$ is the polar angle of the spherical coordinates of $\mathbf{y}$ and we have 
$|\mathbf{x}-\mathbf{y}|^2=(r-\cos\theta)^2+\sin^2\theta$. The formula can be analytically extended in $r$ to the open unit disk in $\C$ to get for  $|z|<1$
\begin{equation}\label{lexp}
 F(z,\theta)
 =
 \sum_{\ell=0}^\infty
 z^\ell P_\ell(\cos\theta)
 \quad\text{with}\quad
 F(z,\theta)=
 \frac{1}{\sqrt{(z-\cos\theta)^2+\sin^2\theta}},
\end{equation}
which is nothing other than the generating function of the Legendre polynomials \cite[\S4.4]{szego}.
The branch of the square root is determined imposing that, as usual, it is real positive for real positive values.
More concretely, we limit the argument to $\alpha\in (-\pi,\pi]$ and take in this range $e^{i\alpha/2}$ to be the square root of $e^{i\alpha}$. 

Let us define for $0<r<1$
\begin{equation}\label{frdef}
 f_r(\theta)=c_rF(r,\theta)
 \qquad\text{with}\quad
 c_r =
 \Big(
 \frac{2\pi}{r}\log\frac{1+r}{1-r}
 \Big)^{-1/2}.
\end{equation}
A calculation proves that this function is normalized as a function on the sphere since
$$
 \int_{S^2}
 |f_r|^2
 =
 2\pi
 \int_0^\pi
 \big|f_r(\theta)\big|^2
 \sin\theta\, d\theta=1.
$$

Note that for each $0<\theta\le\pi$, when $r\to 1^-$, $c_r\to 0$ and $f_r(\theta)\to 0$. This and the previous integral imply that $|f_r|^2$ tends to the Dirac delta on the north pole of the sphere. As a matter of fact, these calculations are very similar to those  showing that the Poisson kernel is an approximation of the identity \cite{rudin}.
\medskip

The function $\big|F(1,\theta)\big|^2\sin\theta$ defined above has a singularity $\sim \theta^{-1}$ when $\theta\to 0^+$ and this singularity moves to $\alpha$ behaving as $\sim (2|\theta-\alpha|)^{-1}$ when we consider $\big|F(e^{i\alpha},\theta)\big|^2\sin\theta$ with $0<\alpha<\pi$. In Figure~\ref{fig:2} there are some plots showing this singular behavior.

\begin{figure}[H]
 \centering
   \begin{tabular}{c}
   \includegraphics[height=100.1823pt]{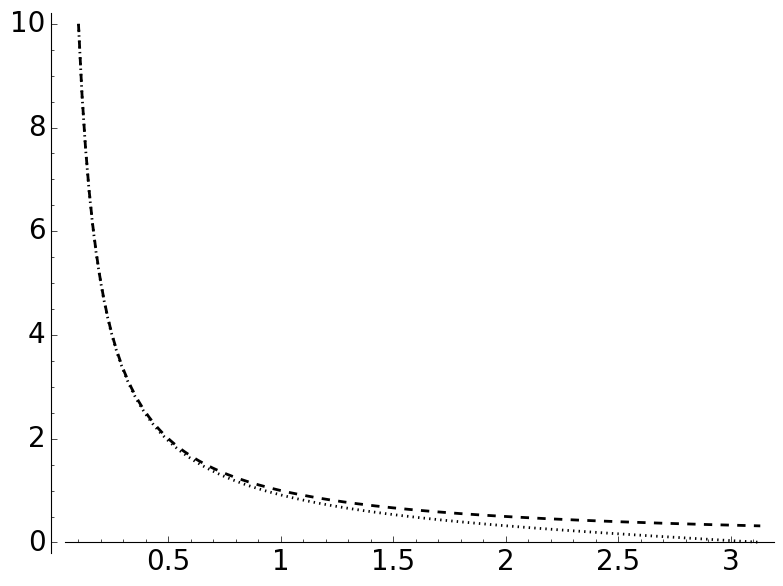}
   \\
   (a)
  \end{tabular}
  \hspace{-10.3049pt}
   \begin{tabular}{c}
   \includegraphics[height=100.1823pt]{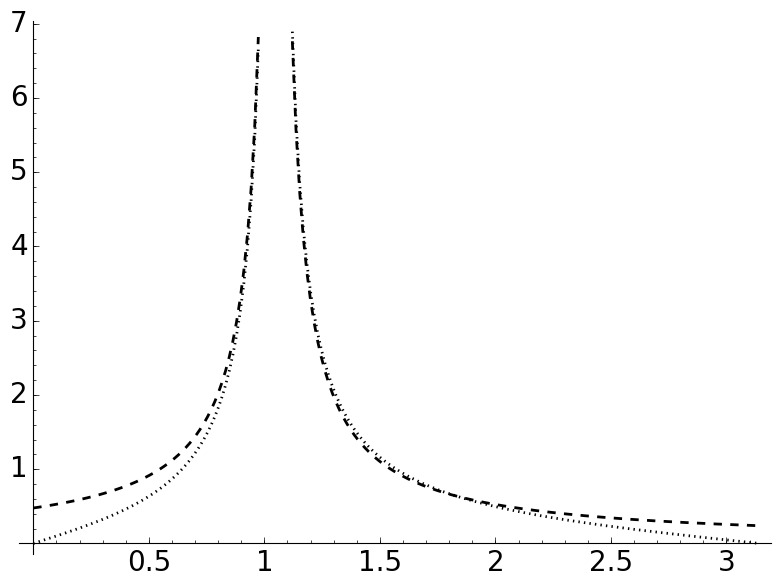}
   \\
   (b)
  \end{tabular}
  \hspace{-10.3049pt}
    \begin{tabular}{c}
   \includegraphics[height=100.1823pt]{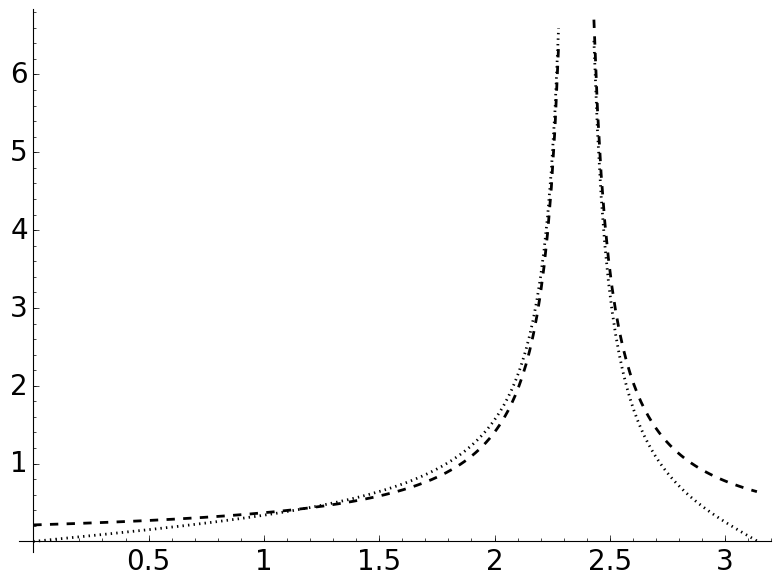}
   \\
   (c)
  \end{tabular}
  \caption{The dashed line corresponds to $\big|F(e^{i\alpha},\theta)\big|^2\sin\theta$ and the dotted line to the approximation of the singularity $s(\theta)$. (a) $\alpha=0$, $s(\theta)=\theta^{-1}$; 
  (b) $\alpha=\frac{\pi}3$, $s(\theta)=\frac{1}{2} {\big|\theta-\frac{\pi}3\big|^{-1}}$; 
  (c) $\alpha=\frac{3\pi}4$, $s(\theta)=\frac{1}{2}{\big|\theta-\frac{3\pi}4\big|^{-1}}$. }
  \label{fig:2}
\end{figure}

These singularities will play a role to study the peaks of the Talbot carpet in the quantum setting.
Note that $\big|F(e^{i\alpha},\theta)\big|=\big|F(e^{-i\alpha},\theta)\big|$. This implies that we can extend the range to $-\pi<\alpha<\pi$ by just changing $\alpha$ by $|\alpha|$. On the other hand, by the symmetry $F(1,\theta)=F(-1,\pi-\theta)$, we can also include the case $\alpha=\pi$.

One can ask what is the aspect of the Talbot carpet for the quantum Talbot effect on the sphere. Figure~\ref{fig:3} corresponds to the density plot of the square root of $|\Psi(\theta,t)|^2\sin\theta$ for $r=0.95$ in our choice of $f_r$.
The vertical axis is $\theta\in [0,\pi]$ and the horizontal axis $t\in [0,\pi]$. The resolution is of $2^{10}$ points on each axis. The computation of $\Psi$ was done through the first 1000 terms of a series expansion. Namely, \eqref{ssolr} with $\ell<1000$.

\begin{figure}[H]
 \centering
   \begin{tabular}{c}
   \includegraphics[height=182.23pt]{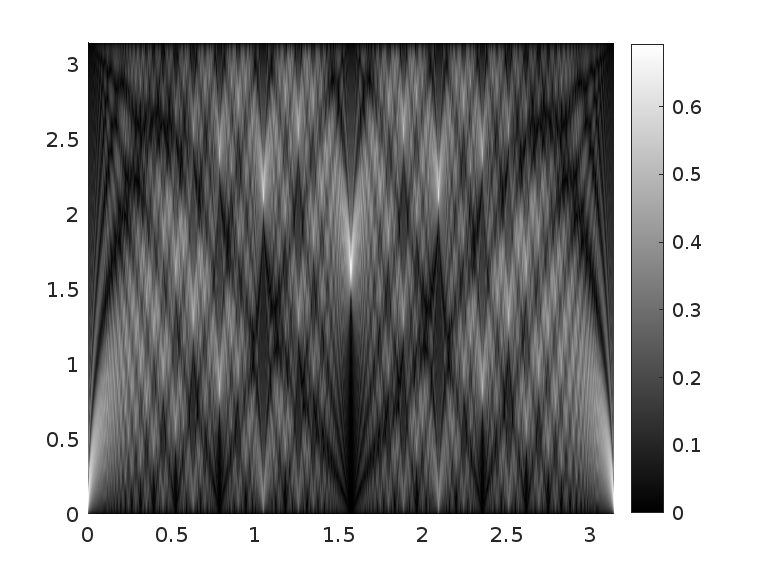}
  \end{tabular}
  \caption{Quantum Talbot carpet for the sphere.}
  \label{fig:3}
\end{figure}

\section{The generalized quadratic Gauss sums}

Before going about the analysis of the Talbot effect in detail, it is  convenient to review some formulas that will be helpful for computing the wave function at special times.
Consider three numbers $a,b,q\in\Z$ with $q>0$ and $a$ coprime. Corresponding to these numbers we introduce the generalized quadratic Gauss sum
\begin{equation}\label{Gdef}
 G(a,b;q)
 =
 \sum_{n=0}^{q-1}
 e
 \Big(
 \frac{an^2+bn}{q}
 \Big).
\end{equation}
As we will show below, the set of values for which it vanishes admits a simple characterization.

\

\emph{Proposition 1:} The quadratic Gauss sum
\begin{equation}\label{gvan}
 G(a,b;q)=0
 \quad
 \text{if and only if}
 \quad
 4\mid 2(b+1)+q.
\end{equation}
Here, the notation $d\mid A$ means $d$ divides $A$.

\

\emph{Proof:} The proof of \eqref{gvan} can be divided in two steps. First, one has to prove that
\begin{equation}\label{Gsq}
 \big|
 G(a,b;q)
 \big|^2
 =
 \left\{
 \begin{array}{ll}
  q &\text{if $q$ is odd,}
  \\
  q\Big(1+e\big(\frac{aq/2+b}{2}\big)\Big)  &\text{if $q$ is even.}
 \end{array}\right.
\end{equation}
The last identity is well known in the context of number theory, but it may be instructive to review its proof below.
By assuming that this identity is already proven, then \eqref{gvan} follows from here because the only way to cancel the latter big parenthesis is to have $aq/2+b$ odd, which is equivalent to $q/2+b$ odd, because $a$ and $q$ are coprime, and $2(b+1)+q=2(q/2+b+1)$ is multiple of $4$ if and only if $q$ is even and $q/2+b$ odd.

The last discussion shows that the only remaining step is to deduce the identity \eqref{Gsq}. For this purpose, write the sum as
$$
 \big|G(a,b;q)\big|^2
 =
 \sum_{m=0}^{q-1}
 e
 \Big(
 \frac{-am^2-bm}{q}
 \Big)
 \sum_{n=0}^{q-1}
 e
 \Big(
 \frac{an^2+bn}{q}
 \Big).
$$
The inner sum is invariant by translations $n\mapsto n+k$, then we can safely change $n$ by $n+m$ and switch later the order of summation to get
$$
 \sum_{m=0}^{q-1}
 \sum_{n=0}^{q-1}
 e
 \Big(
 \frac{a(2mn+n^2)+bn}{q}
 \Big)
 =
 \sum_{n=0}^{q-1}
 e
 \Big(
 \frac{an^2+bn}{q}
 \Big)
 \sum_{m=0}^{q-1}
 e
 \Big(
 \frac{2nam}{q}
 \Big).
$$
If $q\nmid 2n$ then $2na/q\not\in\Z$ and the innermost sum is zero. If $q$ is odd, the divisibility only occurs in the case $n=0$ and $\big|G(a,b;q)\big|^2=q$ is deduced.
If $q$ is even the divisibility also holds for $n=q/2$ and then, in this case,
$$
 \big|
 G(a,b;q)
 \big|^2
 =
  q+q e\Big(\frac{aq^2/4+bq/2}{q}\Big),
$$
giving the result (Q. E. D).

\

The formula \eqref{Gsq} gives of course information about the modulus of the Gaussian sums $G(a,b;q)$, however it does not says anything
about their phase. This quantity depends on some deeper arithmetic properties linked to the quadratic reciprocity (\cite[\S3]{FiJuKo}, cf. \cite[\S9.10]{apostol}).
In a next section it will be important to know its variation in terms of $b$. 
Here we give a statement enough for our purposes. First, we introduce the notation $a_*$ to mean the inverse modulo~$q$ i.e., $a_*\in\Z$ and $q$ divides $a_*a-1$. Of course, the existence of $a_*$ requires $a$ and $q$ to be coprime. In particular
$(4a)_*$ does not make sense if $q$ is even (and it is $4_*a_*$ if $q$ is odd). Let us redefine $(4a)_*$ as~$a_*/4$ when $q$ is even. In this way, we have
\begin{equation}\label{redef}
 (4a)_*4a-1\in\Z
 \qquad\text{and}\qquad
 q\mid (4a)_*4a-1
\end{equation}
irrespective of the parity of $q$.

After this notation, we claim that there exists $G_{a,q}\ne 0$ not depending on $b$ such that
\begin{equation}\label{Gphase}
 e
 \Big(
 \frac{(4a)_*b^2}{q}
 \Big)
 G(a,b;q)
 =
 \left\{
 \begin{array}{ll}
  0&\text{if } 4\mid 2b+q,
  \\
  G_{a,q}&\text{if } 4\nmid 2b+q.
 \end{array}
 \right.
\end{equation}

The first case
is covered by \eqref{gvan} then we assume $4\nmid 2b+q+2$. If $q$ is odd, \eqref{Gphase}
comes from completing squares modulo~$q$. Note that in this case $2_*\in\Z$ and we have
$$
 an^2+bn
 \equiv
 a(n^2+2a_*2_*bn)
 \equiv 
 a(n+2_*a_*b)^2-4_*a_*b^2
 \pmod{q}.
$$
The definition \eqref{Gdef} is invariant by any integral translation $n\mapsto n+k$ and taking $k=2_*a_*b$ we have the result for $2\nmid q$ with 
$G_{a,q}=G(a,0;q)$. 

On the other hand, if $2\mid q$ and $4\nmid 2b+q+2$, necessarily $b+q/2$ is even. Write in this case
$$
 an^2+bn
 \equiv 
 a\Big(n+a_*\frac{b+q/2}{2}
 \Big)^2
 -
 \frac q2 n
 -
 a_*
 \Big(\frac{b+q/2}{2} \Big)^2 
 \pmod{q}.
$$
The last expression can be rearranged as 
$$
 a
 \Big(n+a_*\frac{b+q/2}{2}
 \Big)^2
 -
 \frac q2 
 \Big(n+a_*\frac{b+q/2}{2}
 \Big)
 +
 \frac{a_*q^2}{16}
 -
 \frac{a_*b^2}{4}.
$$
Applying the translation $n\mapsto n+a_*(b+q/2)/2$ in \eqref{Gdef}, we get the result with  
$G_{a,q}=e(a_*q/16)G(a,-q/2;q)$.

\section{A finite form of the solution}
After this digression about Gauss sums, let us consider the task of solving the Schr\"odinger equation \eqref{pde}.
Substituting $a_\ell =c_r r^l$ in \eqref{ssol}, by \eqref{frdef} and \eqref{lexp}, we have that the solution of \eqref{pde} is given by
\begin{equation}\label{ssolr}
  \Psi(\theta,t)
 =
 c_r
 \sum_{\ell=0}^\infty
 r^\ell P_\ell(\cos\theta)e^{-i\ell(\ell+1)t}.
\end{equation}
The sum does not admit for general $t$ a closed expression (actually it is related to integral transforms of elliptic functions) but it can be shown that it does for fractional times $t=2\pi a/q$ with $a/q$ an irreducible fraction, by employing
the properties of the Gauss sums described in the previous section.
We noticed before the time periodicity $\Psi(\theta,t)=\Psi(\theta,t+\pi)$ then we can restrict ourselves to the case $0\leq a/q<1/2$.

For $\ell$ modulo $q$ the map
$\ell\mapsto e\big(\ell(\ell+1)/q\big)$
is well defined and consequently it can be expanded into a finite Fourier series
$$
 e\Big(\frac{\ell(\ell+1)}q\Big)
 =
 \sum_{n=0}^{q-1}
 a_n
 e\Big(-\frac{ln}q\Big).
$$
The coefficients $a_n$ are given by the discrete Fourier transform (DWT) and they become a generalized quadratic Gauss sum, except for a normalizing factor,
$$
 a_n=
 \frac 1q
 \sum_{\ell=0}^{q-1}
 e\Big(\frac{\ell(\ell+1)a+n\ell}q\Big)
 =
 \frac 1 q
 G(a,a+n; q).
$$
Hence
$$
 \Psi\bigg(\theta, \frac{2\pi a}{q}\bigg)
 =
 \frac{c_r}{q}
 \sum_{\ell=0}^\infty
 \sum_{n=0}^{q-1}
 G(-a,-a-n;q)
 r^\ell e\Big(\frac{\ell n}{q}\Big) P_\ell(\cos\theta).
$$
By \eqref{lexp}, it admits a completely explicit expression
\begin{equation}\label{rattsol}
 \Psi\bigg(\theta, \frac{2\pi a}{q}\bigg)
 =
 \frac{c_r}{q}
 \sum_{-q/2< n\le q/2}
 G(-a,-a-n;q)
 F\bigg(re\bigg(\frac{n}{q}\bigg), \theta\bigg).
\end{equation}
The change in the range of $n$ is harmless due to the $q$-periodicity and it will be convenient later for some calculations.
Disregarding the normalizing factor $c_r$,
this formula makes sense if we take formally $r=1$. In fact, 
$c_rG(-a,-a-n;q)F\big(re(n/q\big)$
can be written explicitly as
 \begin{equation}\label{rattsol2}
\frac{ G(-a,-a-n;q)\;e^{i\alpha_n}}{\sqrt{2\pi\bigg[\frac{(r^2-1)^2}{r}+4 r(\cos^2\theta+\cos^2\frac{2\pi n}{q})-4(r^2+1)\cos\theta \cos\frac{2\pi n}{q}\bigg] \log \frac{1+r}{1-r}}},
\end{equation}
where 
$$
\tan \alpha_n=\frac{r^2 \cos\frac{4\pi n}{q}-2 r \cos\theta \cos\frac{2\pi n}{q}+1}{r^2\sin \frac{4\pi n}{q}-2 r \cos\theta \sin\frac{2\pi n}{q}}
$$
By direct inspection it can be seen that if $\cos\theta=\cos\frac{2\pi n}{q}$ then in the limit $r\to 1^-$ the denominator in the first expression (under the square root) involves an indeterminate form $0\cdot \infty$,
which is solved to give zero. Thus, at these angles the wave function becomes singular. For $\cos\theta\neq \cos\frac{2\pi n}{q}$
the limit $r\to 1^-$ gives zero. This shows the distributional behavior of the resulting function, as the support is located at finite points
defined by $\theta=\pm \frac{2\pi n}{q}$, with weights $G(-a,-a-n;q)$. A more detailed analysis of this behaviour will be done in the following sections.
 
Even though we are interested in the limit $r\to 1^-$, it is interesting to study the case $0<r<1$. By the regularity of the solution, we could evaluate $\Psi(\theta,t)$ at any $t$ approximating $t/(2\pi)$ by rationals.
This requires some arithmetic considerations. Given $t/(2\pi)\not\in\Q$, it is known that there exist infinitely many irreducible fractions such that
$$
 \epsilon=q^2
 \Big|
 \frac{t}{2\pi}-\frac aq
 \Big|
 \qquad\text{verifies}\quad \epsilon<1.
$$
In fact, this can be improved to $\epsilon <1/\sqrt{5}$ and not beyond (for general $t$) according to Hurwitz's theorem and that the best approximations are given by the convergents in the continued fraction \cite[\S7.9]{MiTa}. By the mean value theorem
\begin{equation}\label{difri}
 \Psi(\theta,t)
 -
 \Psi\bigg(\theta, \frac{2\pi a}{q}\bigg)
 =
 O
 \big(
 \epsilon q^{-2}|\Psi_t(\theta,2\pi\xi)|
 \big)
\end{equation}
for some $\xi$ between $t/(2\pi)$ and $a/q$, in particular $|\xi-a/q|<1/q^2$. 
Thus, the error requires an estimation of $\Psi_t(\theta,2\pi\xi)$. This can be achieved by using the standard integral representation of the Legendre polynomials \cite[8.913]{GrRy}, which gives in \eqref{ssolr} the following
$$
 \Psi_t(\theta,2\pi\xi)
 =
 -c_r\frac{i\sqrt{2}}{\pi}
 \int_\theta^\pi
 \frac{\sum_{\ell=0}^\infty
 \ell(\ell+1)r^\ell \sin((\ell+1/2)u)e(-\xi\ell^2-\xi\ell)
 }{\sqrt{\cos\theta-\cos u}}
 \,du.
$$
The terms $\ell(\ell+1)r^\ell$ grow until $\ell$ is of order $(1-r)^{-1}$ and decrease exponentially later due to the effect of $r^\ell$ (recall that $0<r<1$ and we have in mind $r$ close to~$1$). Then the main contribution to the sum comes from the terms with $\ell$ comparable to $(1-r)^{-1}$ for which $\ell(\ell+1)r^\ell$ is comparable to $(1-r)^{-2}$. 
In this context, by use of the classical work \cite{HaLi} it can be deduced (see \cite[Th.\,6]{FiJuKo} for a clear statement)
\begin{equation}\label{HLb}
 \sum_{n=M}^{M+N}
 e(\xi n^2+x n)=
 O\Big(
 \frac{N}{\sqrt{q}}
 +
 \sqrt{q}
 \Big)
 \qquad\text{for any}\quad x\in\R.
\end{equation}
Of course, if $q>N^2$, the trivial bound $O(N)$ is better.
The application of this formula shows, by partial summation, that the sum inside the integral is of the order
\begin{equation}\label{erra1}
 O
 \Big(
 (1-r)^{-2}
 \Big(
 \frac{(1-r)^{-2}}{\sqrt{q}}+\sqrt{q}
 \Big)
 \Big)
 \qquad\text{if}\quad q\le (1-r)^{-2}
\end{equation}
and
\begin{equation}\label{erra2}
 O\big(
 (1-r)^{-3}
 \big)
 \qquad\text{if}\quad q\ge (1-r)^{-2}.
\end{equation}
On the other hand, it is not difficult to prove, by taking into account that $\cos\theta-\cos u>C(u^2-\theta^2)$ for $0\le \theta\le u\le \pi/4$, that
\begin{equation}\label{erra3}
 \int_\theta^\pi
 \frac{du}{\sqrt{\cos\theta-\cos u}}
 =
 O\big(|\log\theta|\big).
\end{equation}
The substitution of \eqref{erra1}-\eqref{erra3} into \eqref{difri} leads to the estimation
\begin{equation}\label{irrat}
 \Psi(\theta,t)
 =\Psi(\theta, \frac{2\pi a}{q})+O(E)
\end{equation}
with
$$
 E=\epsilon c_r|\log\theta|\frac{ q^{-1/2}
 +
 \min\big((1-r)\sqrt{q},1\big)
 }{(1-r)^{3}q^{2}}.
$$
Probably the $\log\theta$ factor can be avoided with a more careful analysis because $|P_\ell(\cos\theta)|\le 1$ cannot be achieved with our integral estimation.

To illustrate the situation we plot $\big|\Psi(\theta,t)\big|^2\sin\theta$ in Figure~\ref{fig:4}. 
For $t =2\pi/\sqrt{14}$ we get the first plot.
If we approximate $t/2\pi=1/\sqrt{14}$ by $4/15$, which is a convergent in its continued fraction of $1/\sqrt{14}$, we get the second plot.

\begin{figure}[H]
 \centering
   \begin{tabular}{c}
   \includegraphics[height=118.9722pt]{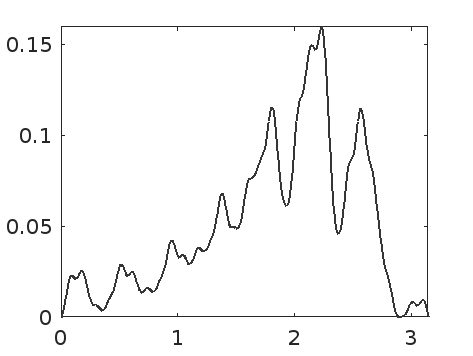}
   \\
   (a)
  \end{tabular}
  \quad
   \begin{tabular}{c}
   \includegraphics[height=118.9722pt]{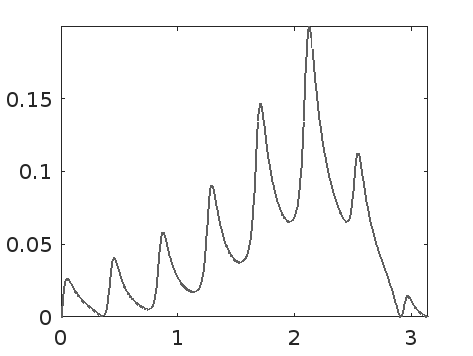}
   \\
   (b)
  \end{tabular}
  \caption{Graph of $\big|\Psi(\theta,t)\big|^2\sin\theta$ for (a) $t=2\pi/\sqrt{14}$ and (b) $t=8\pi/15$.}
  \label{fig:4}
\end{figure}

Clearly the approximation misses the details and it suggests to take the next two convergents $27/101$ and $31/116$, as shown in Figure~\ref{fig:5}.
The last one is barely distinguishable from the original.

\begin{figure}[H]
 \centering
   \begin{tabular}{c}
   \includegraphics[height=118.9722pt]{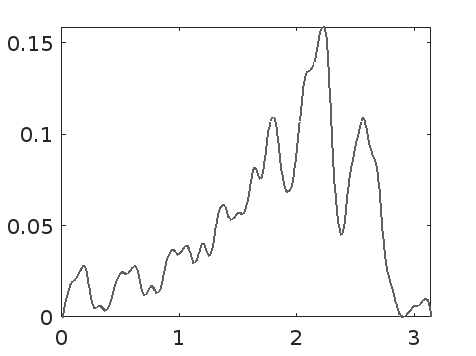}
   \\
   (a)
  \end{tabular}
  \quad
   \begin{tabular}{c}
   \includegraphics[height=118.9722pt]{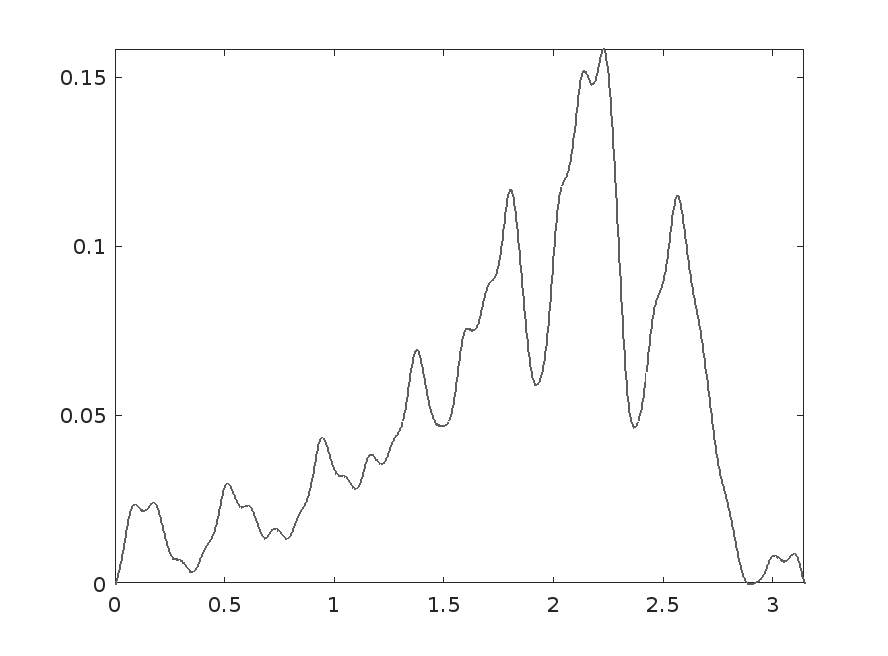}
   \\
   (b)
  \end{tabular}
  \caption{Graph of $\big|\Psi(\theta,t)\big|^2\sin\theta$ for (a) $t=54\pi/101$ and (b) $t=31\pi/58$.}
  \label{fig:5}
\end{figure}

\section{Analysis of the singularities}
We have suggested below (\ref{rattsol2}) that the limit $r\to 1^-$ of the solution is singular.
We would like here to describe this singularity in a more explicit form.
After \eqref{rattsol}, we
\begin{equation}\label{saq}
 S_{a/q}(\theta)
 =
  \sum_{-\frac{q}{2}< n\le \frac{q}{2}}
 G(-a,-a-n;q)
F\Big(e\big(\frac{n}{q}\big),\theta\Big)
\end{equation}
captures the behaviour of the wave function when $r\to 1^-$ at a fixed time $t=2\pi a/q\in [0,\pi]$ with $a/q$ an irreducible fraction. We are going to study the singularities of this function for $\theta\in [0,\pi]$.

Note that $F(e^{i\alpha},\theta)$ is singular exactly at $\theta=|\alpha|$ for $\alpha\in (-\pi, \pi]$ then the only possible singularities occur at $\theta=2\pi k/q$ with $0\le k\le q/2$.
We want to show
\begin{equation}\label{gsing}
 \lim_{\theta\to \frac{2\pi k}{q}^+}|S_{a/q}(\theta)|
 =
 \infty
 ,\quad
 \exists \lim_{\theta\to \frac{2\pi k}{q}^-}S_{a/q}(\theta)
 \qquad\text{if}\quad
 G(a,a+k;q)\ne 0
\end{equation}
and
\begin{equation}\label{ngsing}
 \exists\lim_{\theta\to\frac{2\pi k}{q}}S_{a/q}(\theta)
 \qquad\text{if}\quad
 G(a,a+k;q)= 0.
\end{equation}
Here $\exists$ means that the limit exists and is finite. Of course, one of the lateral limits does not make sense in the extreme points $0$ and $\pi$.

\medskip

A preliminary remark is that $F\big( e(n/q),\theta\big)$ has a not so difficult polar form
\begin{equation}\label{RofF}
 F\Big(e\big(\frac{n}{q}\big),\theta\Big)=
 Re^{iA}
 \qquad\text{with}\quad
 R\big(\frac{n}{q},\theta\big)=
 2^{-1/2}\Big|\cos\frac{2\pi n}{q}-\cos\theta\Big|^{-1/2}
\end{equation}
and
\begin{equation}\label{AofF}
 A(\frac{n}{q},\theta)=
 \left\{
 \begin{array}{ll}
  -\frac{\pi n}{q}&\text{if }\frac{2\pi|n|}{q}<\theta<\pi,
  \\[7pt]
  \frac{\text{\scriptsize\rm sgn}(n)\pi}{2}-\frac{\pi n}{q} &\text{if }0\le\theta<\frac{2\pi|n|}{q}.
 \end{array}\right.
\end{equation}
These formulas may be found by taking limits in \eqref{rattsol2}. A more practical approach is the following.
First note that
$\big(e(\frac{n}{q})-\cos\theta\big)^2+1-\cos^2\theta$ is a first order polynomial in $\cos\theta$ which vanishes for $\theta=2\pi n/q$, then it must be divisible by $\cos(2\pi n/q)-\cos\theta$, resulting
$$
 \bigg(e\bigg(\frac{n}{q}\bigg)-\cos\theta\bigg)^2+\sin^2\theta
 =
 2e\bigg(\frac{n}{q}\bigg)\bigg(
 \cos\bigg(\frac{2\pi n}{q}\bigg)-\cos\theta
 \bigg).
$$
From here, the formula for $R$ becomes obvious, recalling the definition of~$F$.
If $2\pi|n|/q<\theta<\pi$
the latter big parenthesis is positive and then
$F\big( e(n/q),\theta\big)/R=e(-n/2q)$
giving the first case in \eqref{AofF}.
On the other hand, if
$0\le\theta<2\pi|n|/q$,
it is negative and we recover  the positivity introducing a factor $e(\pm 1/2)$. To keep the argument of $e(\pm 1/2+n/q)$ in the range $(-\pi,\pi]$, we choose the minus sign if and only if $n$ is positive. Then
$F\big( e(n/q),\theta\big)/R=e\big(-\frac 12(\textrm{sgn}(n)/2+n/q)\big)$
and the second case of the formula is deduced.
\medskip

Now we prove \eqref{gsing} and \eqref{ngsing}.
Note that
$G(-a,-a-k;q)=0$
if and only if
$G(a,a+k;q)=0$
because both are complex conjugates.
Let us discuss first the extreme cases. If $\theta\to 0^+$ then $F(1,\theta)\sim \theta^{-1}$ and we have in \eqref{saq}
$$
 S_{a/q}\sim G(-a,-a;q)\theta^{-1} +\text{constant terms},
$$
giving the result. When $\theta\to \pi^-$ the only possible singular term in \eqref{saq} is that with $n=q/2$ that only exists if $2\mid q$, but $G(-a,-a-q/2;q)=0$ by \eqref{gvan} and the singularity is suppressed.

Consider  now the remaining cases $\theta\to 2\pi k/q$ with $0<k<q/2$. The possible singularity comes from
\begin{equation}\label{singp}
 G(-a,-a+k; q)
 F\big( e(-k/q),\theta\big)
 +
 G(-a,-a-k; q)
 F\big( e(k/q),\theta\big).
\end{equation}
If $G(-a,-a-k; q)=0$ then $G(-a,-a+k; q)=0$ by \eqref{gvan} because $-a-k$ and $-a+k$ have the same parity. Then there is no singularity when $G(a,a+k; q)=0$ and we get \eqref{ngsing}.

If $G(a,a+k; q)\ne 0$,
applying \eqref{Gphase}, \eqref{RofF} and \eqref{AofF} to \eqref{singp}, we see that \eqref{gsing} follows if we prove
$$
 e
 \Big(
 \frac{(4a)_*(a-k)^2}{q}
 +
 \frac{n}{2q}
 \Big)
 +
 e
 \Big(
 \frac{(4a)_*(a+k)^2}{q}
 -
 \frac{n}{2q}
 \Big)
 \ne 0
$$
and
\begin{equation}\label{leftv}
 e
 \Big(
 \frac{(4a)_*(a-k)^2}{q}
 +
 \frac{n}{2q}
 -\frac 14
 \Big)
 +
 e
 \Big(
 \frac{(4a)_*(a+k)^2}{q}
 -
 \frac{n}{2q}
 +\frac 14
 \Big)
 = 0
\end{equation}
because, except for a nonzero constant, these are the results of dividing \eqref{singp} by $R(k/q,\theta)$ when $\theta$ is, respectively, to the right and to the left of $2\pi k/q$. It is enough to prove
$$
 \frac{(4a)_*(a-k)^2}{q}
 +
 \frac{n}{2q}
 -
 \Big(
 \frac{(4a)_*(a+k)^2}{q}
 -
 \frac{n}{2q}
 \Big)
 \in\Z
$$
because it means that the complex exponentials are equal in the first case and have opposite sign in the second case as a result of the arguments in \eqref{leftv} differing in a half-integer. Opening the squares, we have
$$
 -\frac{(4a)_*4ak}{q}
 +\frac kq
 =
 -\frac{\big((4a)_*4a-1\big)k}{q}
$$
and this is an integer by \eqref{redef}.

\section{The ``valleys of the shadows''}
The work \cite{oskolkov} formalizes the mathematical study of the evolution under the Schr\"odinger equation on the circle (equivalently, on the real line under periodic conditions) of wave functions~$\Psi$ such that 
the probability density 
$|\Psi|^2$ tends to the periodic Dirac delta $\delta_p(x)=\sum_n \delta(x-n)$ at the initial time $t=0$.  
In that work, the author considers generic periodic $\sqrt{\delta}$ families, which are functions $f_\epsilon(x)$ parametrized by a real quantity $\epsilon$ in such a way that
$$
\lim_{\epsilon\to 0}\int_0^1 |f_\epsilon(x)|^2 g(x)dx=g(0),
$$
for every function $g(x)$ continuous in the interval $[0,1]$. Clearly, this condition states that $f(x)=\lim_{\epsilon\to 0}f_\epsilon(x)$ imitates the behaviour of the square root of a Dirac delta distribution. An example of such is $$f_\epsilon(x)=(2\pi \epsilon^2)^{-\frac{1}{4}}e^{-\frac{x^2}{2\epsilon}},$$
however the results of that reference do not rely in any particular family. The only distinction is that the family can be even  
or odd, according $f_\epsilon(-x)$ equals $f_\epsilon(x)$ or $-f_\epsilon(x)$, respectively, and the even families are denoted $\sqrt{\delta_+}$ and the odd ones $\sqrt{\delta_-}$. In general, a given distribution $\rho(f)$ is the weak limit of $\rho(f_\epsilon)$
if for every continuous and compactly supported $g$ in the circle $\mathbb{T}$
$$
\lim_{\epsilon\to 0}\int_{\mathbb{T}} \rho(f_\epsilon) g\, d\mu=\int_{\mathbb{T}} \rho(f) g\, d\mu,
$$
with $\mu$ the standard Lebesgue measure on the circle. For the particular case of the Schr\"odinger equation, the distribution $\rho(f)$ represents the probability density $|\Psi|^2$ evolving in time with initial data $f$.

In the terms of the quantities defined above, by assuming that the limit $f(x)$ does not depend on the choice of the family,  one of the main results of the reference \cite{oskolkov}
is the following

\

\emph{Proposition 2:} 
Consider the straight lines 
$$
 L_{a, b}=\{(x,t) \in \R^2\;|\;\; x+a t=b\},\qquad L^T_{a, b}=\{(x,t) \in \R^2\;|\;\; ax+t=b\},
$$
and let $\rho(\sqrt{\delta}, C)$ denote the restriction of the density to a set $C\subset\R^2$.
Given $N$ and $M$ integers, then 
$$
\rho(\sqrt{\delta_\pm}, L_{N, \frac{M}{2}})=1\pm (-1)^{MN}.
$$
Furthermore if $N$ is rational and not integer then
$$
\rho(\sqrt{\delta}, L_{N, \xi})=\rho(\sqrt{\delta}, L^T_{0, \tau})=\rho(\sqrt{\delta}, \R^2=1,
$$
for $\xi$ real and $\tau$ real irrational.

\

Some comments about this proposition are in order. The identities given there should be interpreted in weak sense \cite{oskolkov}. The formula for $\rho(\sqrt{\delta_\pm}, L_{N, \frac{M}{2}})$ shows that if  the family is even, then for $M$ and $N$ both odd or even, the density vanishes. For odd families instead, one integer should be even and the other odd. The corresponding lines $L_{M, \frac{N}{2}}$ are denominated in 
the terminology of that reference as valley of shadows. The presence of these valleys is a feature that was observed by numerical simulations, see \cite{oskolkov} and reference therein.
 It is important to remark that the position of these valleys depends on the choice of the approximating family. It may be natural to employ an even family for modeling the Dirac delta.

In addition, it is proved \cite[Th.\,1, Lem.\,1]{oskolkov} that the vanishing of the generalized Gauss sum $G(a,b;q)$ with $a/q$ and $b/q$ related to time and position assures that the corresponding point belongs to the valleys of the shadows.

The arguments of \cite{oskolkov} cannot adapted changing the circle by the sphere to cover our case, mainly by two reasons. First, the eigenfunctions for the sphere have nothing to do with the pure complex exponentials and in particular they do not share  their additive properties linked to standard operations with the classic Fourier expansion. On the other hand, in some sense (not explicitly mentioned in \cite{oskolkov}), if the time is a rational multiple of $2\pi$, $\Psi$ tends to a single generalized quadratic Gauss sum when working on the circle, while~\eqref{rattsol} shows that an arbitrarily large number of these sums, depending on the denominator, contribute to $\Psi$ when working on the sphere.

Despite the problem for generalizing the results of \cite{oskolkov} given above, there is a characterization that can be expressed in mathematical terms as follows. First, we  define the \emph{rational valleys of shadows} (meaning the valleys of the shadows at times given by rational multiples of $2\pi$) by the formula
\begin{equation}\label{vall}
 \mathcal{V}=
 \Big\{
 (\theta,t)\in [0,\pi]^2\,:\,
 t=\frac{2\pi a}{{q}},\
 S_{a/q}(\theta)=0
 \Big\}.
\end{equation}
Here, as before, $a/q$ represents an irreducible fraction. 
Keeping in mind the formula~\eqref{rattsol}, $\mathcal{V}$ corresponds to empty zones when $r\to 1^-$. In some sense, it is a quantum counterpart of the shadowed zones that appear in the classical diffraction setting.
We will see later how it is reflected in the numerical calculations.

Taking into account the subtle arithmetic information appearing in the exact evaluation of $G(a,b;q)$ (see \cite[\S3]{FiJuKo}) and that $F(z,\theta)$ does not seem to share any arithmetic significance, it is hard to expect a simple characterization of~$\mathcal{V}$.
Surprisingly, using our study on the singularities, we can state a neat conjecture and to verify it except for a thin set.  First, a proposition is needed.

\

\emph{Proposition 3:} The following set inclusion
\begin{equation}\label{vallinc}
 \mathcal{V}\supset
 \mathcal{V}_0
 \qquad\text{where}\quad
 \mathcal{V}_0
 =
 \bigcup_{{\begin{array}{c}\scriptstyle 0\le \frac{a}{q}\le \frac{1}{2}\\[-3pt] \scriptstyle 4\mid q\end{array}}}
 [0,\frac{2\pi}{q})
 \times 
 \{\frac{2\pi a}{q}\},
\end{equation}
is true.

\

Our conjecture is that $\mathcal{V}=\mathcal{V}_0$, more precisely the following statement.

\

\emph{Conjecture:} Each time slice
\begin{equation}\label{slice}
 \big(\mathcal{V}-\mathcal{V}_0\big)
 \cap \big(
 [0,\pi]
 \times
 \{\frac{2\pi a}{q}\}
 \big)
\end{equation}
contains at most finitely many points.

\

\emph{Proof of proposition 3:} First, take $q$ such that $4\mid q$. Then
$$G(-a,-a-n;q)\ne0,$$ if and only if $n$ is odd, by \eqref{Gphase}, and we can arrange $S_{a/q}(\theta)$ as
$$
 S_{a/q}(\theta)=\sum_{{\begin{array}{c}\scriptstyle k=1\\ \scriptstyle 2\nmid k\end{array}}}^{q/2-1}
 \Big(
 G(-a,-a+k; q)
 F\big( e(-\frac{k}{q}),\theta\big)
 +
 G(-a,-a-k; q)
 F\big( e(\frac{k}{q}),\theta\big)
 \Big).
$$
The function under the sum is \eqref{singp} and we had proved with \eqref{leftv} that it vanishes when $\theta$ is to the left of $2\pi k/q$. As $k\ge 1$ any $0\le\theta<2\pi/q$ cancel all the terms and we obtain \eqref{vallinc} (Q. E. D).

\

The next step is to motivate the conjecture given above. The functions $F\big( e(k/q),\theta\big)$ are real analytic except for the singularities
described in the previous sections. From there it is concluded that $S_{a/q}(\theta)$ is analytic in each open interval $I=(2\pi k_1/q,2\pi k_2/q)$ where $2\pi k_1/q$ and $2\pi k_2/q$ are consecutive singularities of $S_{a/q}$.
By \eqref{gsing} we know that $S_{a/q}$ goes to $\infty$ in the left extreme of $I$, then we can find $\alpha\in I$ such that $S_{a/q}(\theta)\ne 0$ for $\theta\in (2\pi k_1/q,\alpha]$. On the other hand, \eqref{leftv} shows that $S_{a/q}(\theta)$ could be redefined to the right of $2\pi k_2/q$ in an analytic way just setting \eqref{singp} to zero. Then $S_{a/q}$ is analytic in $[\alpha,2\pi k_2/q]$. The non identically zero analytic functions defined on finite closed intervals can only have a finite number of zeros, then the same can be said for $S_{a/q}(\theta)$ in $I$ and the same argument applies for the right half-open extreme interval of the form $(2\pi k_1/q,\pi]$.

Summing up, if we take the union of all of these intervals, we have that $S_{a/q}$ has a finite number of zeros in $[2\pi k_0/q,\pi]$ with $2\pi k_0/q$ the first singularity.
Since \eqref{gvan}, $G(a,a,q)\ne 0$ if and only if $4\nmid q$ and $G(a,a+1,q)\ne 0$ if $4\mid q$ and \eqref{gsing} shows $k_0=0$ in the first case and $k_0=1$ in the second, proving that \eqref{slice} contains finitely many points (conjecturally none). This is basically the motivation for stating that conjecture.
\medskip

Except in symbolic computation environments, it is impossible to distinguish with a computer~$0$ from nearly~$0$. A method to try to approximate the valleys of the shadows is to select atypically small values with certain tolerance.

Figure~\ref{fig:6} displays in black the points in the quantum Talbot carpet of \S2 (recall that $r=0.95$) having values less than a $5\%$ and a $2.5\%$ of the maximum value.
Although we do not obtain very clear pictures, the peaks $[0,2\pi/q)\times \{2\pi a/q\}$ in $\mathcal{V}_0$ with $4\mid q$ for small $q$ are apparent in the bottom part.

\begin{figure}[H]
 \centering
   \begin{tabular}{c}
   \includegraphics[height=122.1712pt]{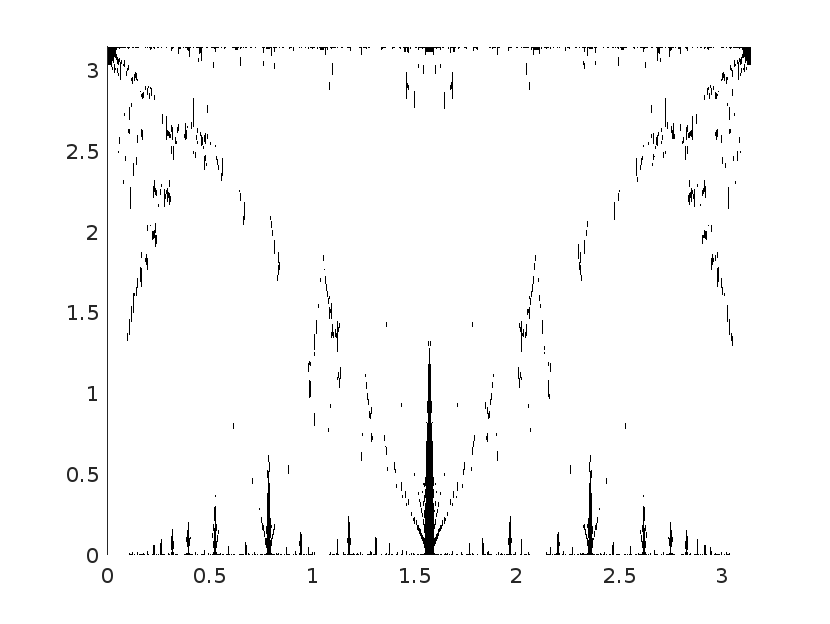}
   \\
   (a)
  \end{tabular}
   \begin{tabular}{c}
   \includegraphics[height=122.1712pt]{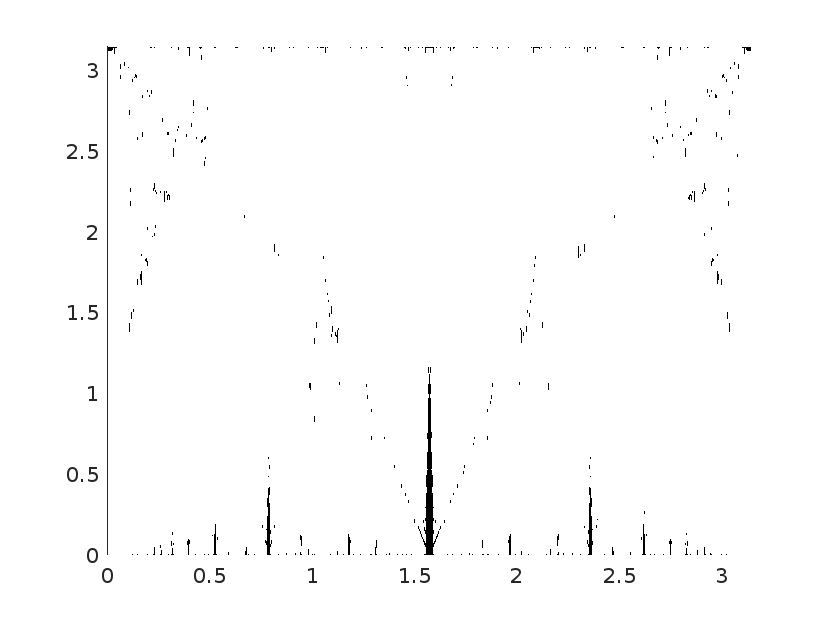}
   \\
   (b)
  \end{tabular}
  \caption{In black, points in the Talbot carpet with values below (a) $5\%$ and (b) $2.5\%$ of the maximum.}
  \label{fig:6}
\end{figure}

\section{The fractional quantum Talbot effect}

As is well known, the study of Gauss sums prove to be fruitful for studying optics, in particular the celebrated Talbot effect about image revivals in a diffraction setting, as described in \cite{BeKl}. This replication translated to the Schr\"odinger context can be interpreted as follows. We know that $G(a,a+k;q)=0$ if and only if $q/2+a+k+1$ is an even integer. If not, \eqref{gsing} assures a singularity of $S_{a/q}(\theta)$ at $\theta=2\pi k/q$. Recalling \eqref{rattsol}, this can rephrased as the following form of the Talbot effect:

\begin{quote}\it
 The singularity of the initial condition of \eqref{pde} reflected as a peak $\sim\theta^{-1}$ at $\theta=0$ when $r\to 1^-$  in $|f_{r}(\theta)|^2c_{r}^2\sin\theta$, reappears  (at different scale) in its solution for $t=2\pi a/q$ at $\theta=2\pi k/q$, with $0\le k<q/2$ exactly when $2(a+k+1)+q$ is not a multiple of $4$.
\end{quote}

We see that both in the optics and the Schr\"odinger case the singularities result in a partial replication of the initial profile.
There are some remarks to make. The factor $c_r$ in \eqref{rattsol} goes to zero and it was not included in \eqref{saq}. It does not kill the singularities because $c_r\sim\big(-2\pi\log(1-r)\big)^{-1/2}$ and
$$
 \big|F(re(n/q),\theta)\big|
 =
 (1-r)^{-1/2}
 \big(1+r^2-2r\cos(2\theta)\big)^{-1/2}
 \qquad\text{for}\quad \theta=\frac{2\pi n}{q}.
$$
One may also ask whether the singularity of $S_{a/q}(\theta)$ at $\theta=0^+$ when $4\nmid q$ disappears when we consider the density
$\big|\Psi(\theta,2\pi a/q)\big|^2\sin\theta$. Of course, it takes the value zero at $\theta=0$ for any $r<1$, but $\big|F(1,\theta)\big|\sim\theta^{-1}$ suggests a behavior like $\theta^{-1}$ to the right (see \S3).

In all the cases, \eqref{gsing} predicts a bias to the right of the singularity that it is reflected in the numerical calculations.

\medskip

When $q$ is odd the condition in the previous statement is always fulfilled and then we have singularities at $\theta = 0,2\pi/q,4\pi/q,\dots, \pi(q-1)/q$.
\smallskip

In Figure~\ref{fig:7} there are some examples of the graph of
$|\Psi(\theta,2\pi a/q)|^2\sin\theta$
for $q$ odd.
Note that increasing the value of $r$ makes the peaks more noticeable.

\begin{figure}[H]
 \centering
   \begin{tabular}{c}
   \includegraphics[height=78.2372pt]{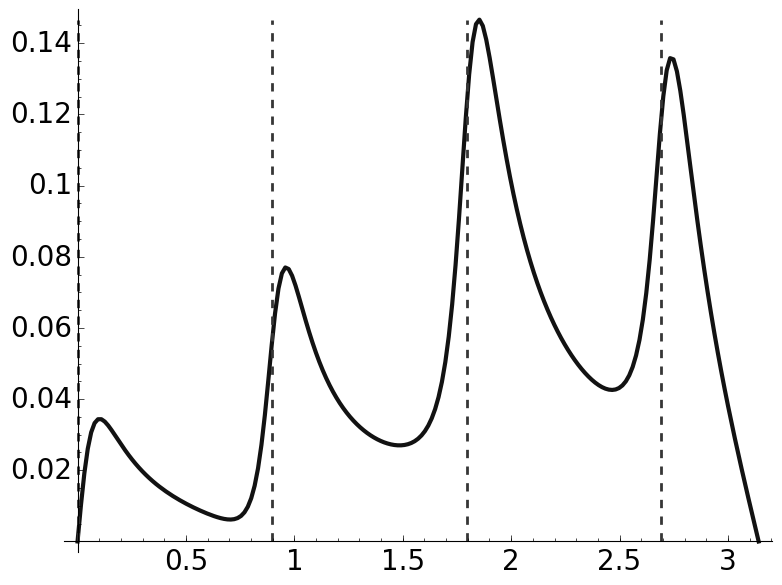}
   \\
   (a)
  \end{tabular}
   \begin{tabular}{c}
   \includegraphics[height=78.2372pt]{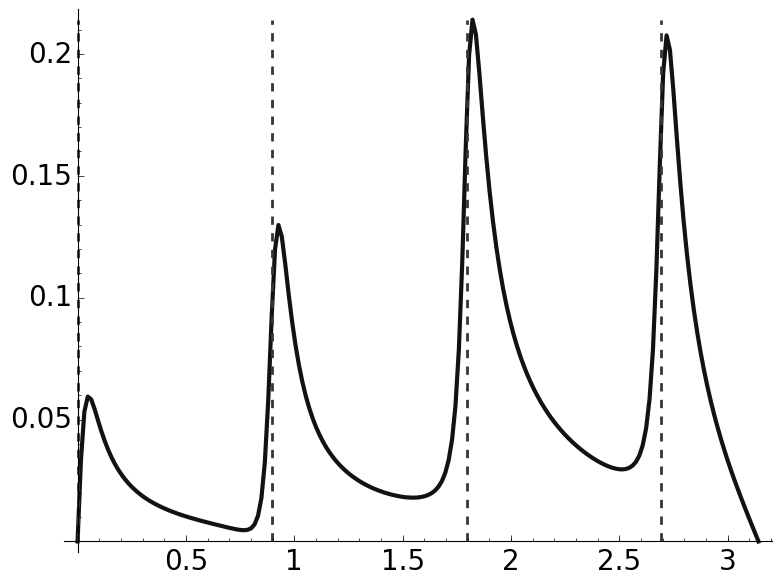}
   \\
   (b)
  \end{tabular}
   \begin{tabular}{c}
   \includegraphics[height=78.2372pt]{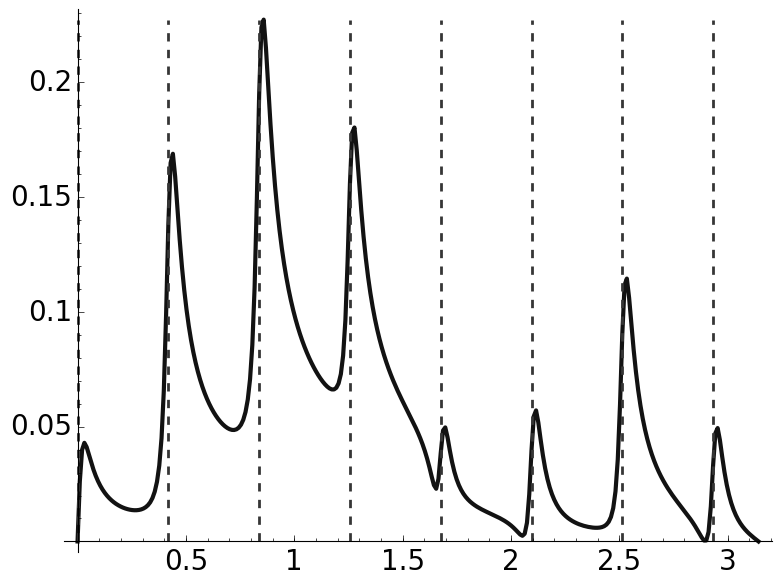}
   \\
   (c)
  \end{tabular}
  \caption{Graph of $|\Psi(\theta,2\pi a/q)|^2\sin\theta$ corresponding to 
  (a) ${a/q=2/7}$, ${r=0.9}$;
  (b) ${a/q=2/7}$, ${r=0.95}$;
  (c) ${a/q=7/15}$, ${r=0.97}$.}
  \label{fig:7}
\end{figure}

When $q$ is even, the condition means that $n$ and $q/2$ have different parity. Hence, if $q$ is a multiple of $4$, the singularities are located exactly at $\theta = 2\pi/q,6\pi/q,\dots, \pi-2\pi/q$ and if $q$ is not, they appear at $0, 4\pi/q,8\pi/q,\dots, \pi-2\pi/q$. The extreme point $\theta=\pi$ is never a singularity, as mentioned in \S6.

Figure~\ref{fig:8} contains some examples for $q$ even.
The last plot illustrates the smoothing of the peaks when $r$ is not close to~$1$.
The flat left part of the first plot is in agreement with our considerations about the valleys of the shadows.

\begin{figure}[H]
 \centering
   \begin{tabular}{c}
   \includegraphics[height=78.2372pt]{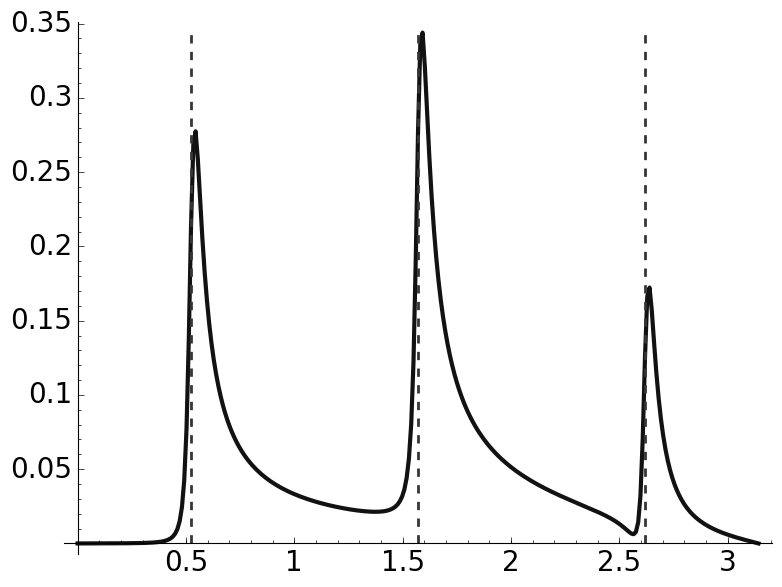}
   \\
   (a)
  \end{tabular}
   \begin{tabular}{c}
   \includegraphics[height=78.2372pt]{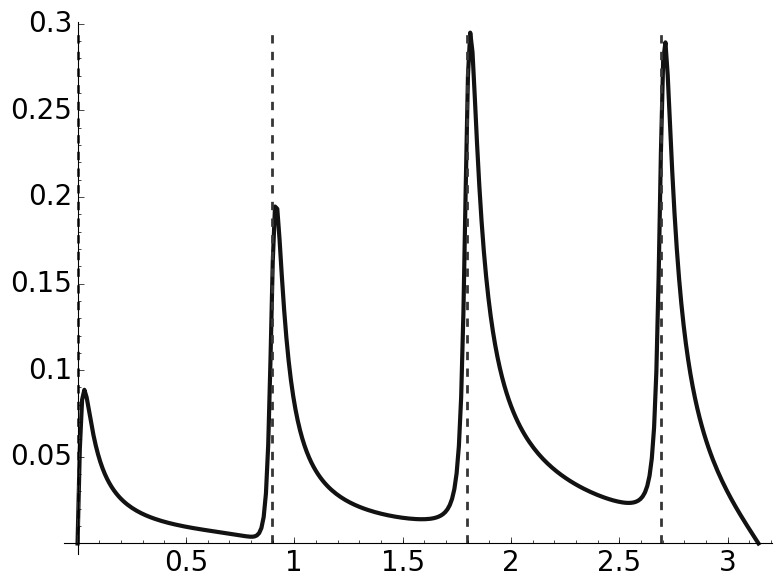}
   \\
   (b)
  \end{tabular}
   \begin{tabular}{c}
   \includegraphics[height=78.2372pt]{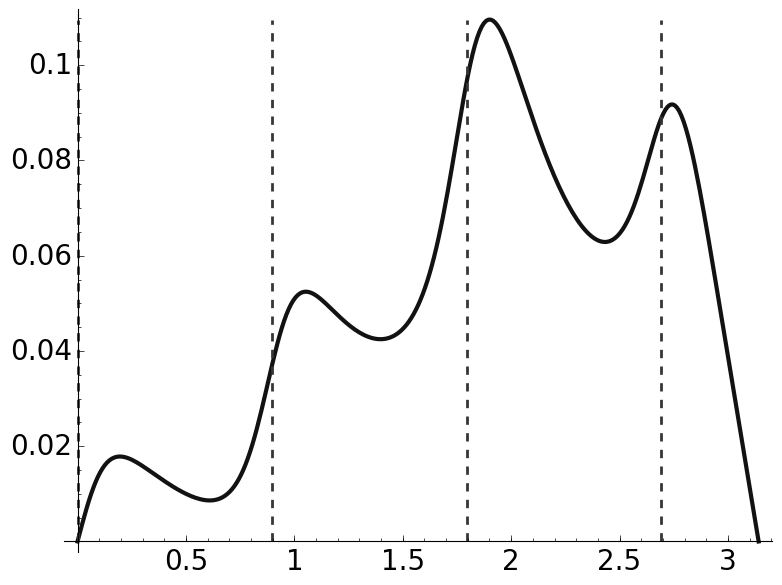}
   \\
   (c)
  \end{tabular}
  \caption{Graph of $|\Psi(\theta,2\pi a/q)|^2\sin\theta$ corresponding to 
  (a) ${a/q=1/12}$, ${r=0.97}$;
  (b) ${a/q=3/14}$, ${r=0.97}$;
  (c) ${a/q=3/14}$, $\mathsf{r=0.8}$.}
  \label{fig:8}
\end{figure}

\section{Conclusions}

In the present work the Schr\"odinger equation for a particle in a two dimensional sphere was studied, by assuming an initial profile which
tends to the Dirac delta in certain limit of the parameters. The structure of singularities at rational time was clarified and it was shown that the density becomes
localized at certain point along the sphere at those specific times. 
It is obtained a set in which the density vanishes and it is conjectured that this set is indeed, with minor variations, the zero set of the density, but a proof is still missing.
It is noticeable that the whole analysis has an arithmetic flavor and number theoretical questions enters in our analysis.
It may be desirable to make a more formal classification of these vanishing regions. In the circle case, the vanishing regions are one dimensional lines. 
In the present case, it is a sort of analogous situation, since these regions are numerable unions of segments.
We leave this issue of a better understanding of this region for a future publication.

\section*{Data availability statement}
All data that support the findings of this study are included within the article (and any supple-
mentary files) and illustrated at the link \url{http://matematicas.uam.es/~fernando.chamizo/dark/d_talbot.html}.

\section*{Acknowledgments}
O.P. S. is grateful to the Universidad Autónoma and to the ICMAT in Madrid, where this work was performed, by their hospitality. The present work is supported by CONICET, Argentina and by the Grant PICT 2020-02181. This project has received funding from the European Union’s Horizon 2020 research and innovation program under the Marie Sk{\l}odowska-Curie grant agreement No 777822.
F. Ch. is partially supported by the PID2020-113350GB-I00 grant of the MICINN (Spain) and
by ``Severo Ochoa Programme for Centres of Excellence in R{\&}D'' (CEX2019-000904-S).

\end{document}